\documentclass[journal=jacsat,manuscript=article]{achemso}
\usepackage[version=3]{mhchem} 

\usepackage[T1]{fontenc}
\usepackage{textcomp,mathcomp}
\usepackage{siunitx}

\author{Liubov Ivzhenko}
\email{ivzhenko@amu.edu.pl}
\affiliation{Institute of Spintronics and Quantum Information, Faculty of Physics and Astronomy, Adam Mickiewicz University, Uniwersytetu Pozna\'nskiego 2, 61-614 Pozna\'n, Poland}

\author{Sergey Polevoy}
\affiliation{Radiospectroscopy dept., O.Ya. Usikov Institute for Radiophysics and Electronics of the NASU, Ak. Proskury St., 12, 61085, Kharkiv, Ukraine}

\author{Sergey Nedukh}
\affiliation{Radiospectroscopy dept., O.Ya. Usikov Institute for Radiophysics and Electronics of the NASU, Ak. Proskury St., 12, 61085, Kharkiv, Ukraine}

\author{Maciej Krawczyk}
\email{krawczyk@amu.edu.pl}
\affiliation{Institute of Spintronics and Quantum Information, Faculty of Physics and Astronomy, Adam Mickiewicz University, Uniwersytetu Pozna\'nskiego 2, 61-614 Pozna\'n, Poland}

\title[An \textsf{achemso} demo]
  {Influence of photon-magnon coupling to enhance spin-wave excitation}  
  
\abbreviations{MW,MSTL,ISRR,SW, SMP}

\begin{document}
\begin{abstract}
   One of the main challenges in magnonics is the efficiency of the conversion of microwave signals into spin waves. This efficiency is low due to the significant mismatch between microwave and spin wave wavelengths in the GHz range $10^{-2}$ m and $10^{-8}$ m, respectively, leading to high energy consumption in magnonic circuits. To address this issue, we propose an approach based on a planar inverse split-ring resonator (ISRR) loaded with a nanometer-thick Py film and exploiting the photon-magnon coupling effect. Our numerical studies show that the ISRR-based antenna achieves more than a fourfold improvement in conversion efficiency compared to a conventional single microstrip transmission line at frequencies and bias magnetic fields around the anti-crossing frequency gap. This has been demonstrated in the weak photon-magnon coupling regime for the nanometer-thin permalloy film with micrometer lateral dimensions. Further optimization of the ISRR can help to achieve the strong coupling regime, making the system potentially useful for quantum technology.
   Our compact and efficient antenna design offers a significant advantage over standard microstrip lines, paving the way for scalable and powerful magnonic circuits for microwave signal processing.
\end{abstract}

\section{Introduction}
Conventional electronic processors face the challenge of further miniaturization and overcoming the existing clock frequency limitation due to Ohmic heating during the passage of an electric current\cite{Horwitz2014,Manipatruni2018}. The actively developing research field of magnonics\cite{Kruglyak_2010,barman2021magnonics} offers an alternative by using spin waves (SWs) instead of electric current to process information\cite{Chumak2022,Wang2023Perspective,Lee2022Reservoir}. The generation of short-wavelength SWs is a necessary step to enable miniaturization of SW-based logic devices, as the device size is limited by the wavelength\cite{mahmoud2020}. Various concepts have already been proposed for SW excitation in ferromagnetic materials. These include voltage modulated anisotropy\cite{Hamalainen2017,Rana2019}, spin currents\cite{Demidov2017,Fulara_2019}, or microwaves (MW) in various configurations\cite{Gruszecki2016,Wintz2016-xc,Yu2016}. The most conventional approach is a single microstrip transmission line (MSTL)~\cite{Mori_2022,Dmitriev1988} or a coplanar waveguide (CPW)~\cite{Goto2019,Sekiguchi2012} as a SW antenna, usually oriented perpendicular to the long axis of a waveguide~\cite{Gro2019,Lisiecki2019}. These antenna configurations have found wide application due to their simplicity in exciting SWs via an alternating magnetic field (\( \mathbf{h} \)-field) generated by the MW current in the metallic strip. 
However, CPW and MSTL antennas are typically designed for far-field MW applications, and they are not inherently optimized for near-field localized MW-SW conversion at a scale of nanometers\cite{Ganguly1975}. 
Attempts have been made to overcome these limitations\cite{Aquino2022,bruckner2025,Erdelyi2025}, for instance, by designing a \(\Omega\) shaped antenna and coupling it to the MW resonator~\cite{Banholzer_2011}. Such an antenna allows SWs to be excited in submicrometer ferromagnetic elements with an out-of-plane MW magnetic field and has been successfully used for advanced magnonic applications, i.e., neuromorphic computing~\cite{Korber2023}. However, it does not fully satisfy the requirements and needs of MW-to-SW conversion efficiency, among others, because the magnetic field distribution obtained in this way is inhomogeneous and for a limited size of ferromagnets. 

We propose to use an alternative approach and exploit photon-magnon coupling for SW excitation. Several types of MW resonators have been tested for photon-magnon coupling. These include cryogenic superconducting resonators, which can have very large quality factors reaching a few thousand. This allows for high cooperativity (e.g., $\approx 30$) even at anti-crossing frequency gap width well below 100 MHz for YIG\cite{Guo2023} and 170 MHz in thin Py stripe\cite{Hou_2019}. At room temperature, MW cavities provide a high concentration of the microwave magnetic field, homogeneous in the particular parts of the cavity, thus relevant for the coupling with the uniform mode of the ferromagnet, often of spherical shape\cite{Tabuchi2014,Zhang2014}. From the point of view of the integration, the planar resonators with ferromagnetic thin films are more promising\cite{Bhoi2017}. These include split-ring resonators\cite{Kaffash_2023} or inverse split-ring resonators (ISRRs)\cite{Kim2024} or spiral resonators\cite{Xiong2024}, for which strong photon-magnon coupling has been demonstrated. In most cases,  the bulk ferromagnetic sample, with linear dimensions comparable to those of the resonator, have been used.  Nonetheless, even in these devices, the position of the ferromagnet with respect to the resonator influences the coupling strength\cite{Macedo2021,Yadav2024,Wagle_2024}, with the common approach being to place the ferromagnet on the side of the feeding line, not of the resonator, if they are on opposite sides\cite{Kim2024,Bhoi2017}. However, these designs do not allow the local enhancement of the near-field distribution of the MW magnetic field of the resonator to be fully exploited. In fact, this prevents exploiting the increased photon-magnon coupling for magnonics or using the advantages of SWs over MWs in signal processing, such as a few orders-of-magnitude shorter wavelength at the same frequency, and a lack of charge carriers. This limits miniaturization and decreasing the energy consumption of MW devices\cite{levchenko2024}.

We show that an ISRR is particularly useful for this purpose, where the metallized narrow strip (anti-gap strip) connecting the internal patch to the surrounding metal patch can receive an oscillating MW current at resonance, fed by a wide microstrip line. The locally generated MW magnetic field couples to the SWs in the ferromagnetic thin metallic film (Py) deposited just above the anti-gap. Importantly, not only the uniform SW mode (fundamental mode) couples, but also modes quantized along the width of the ferromagnet.  The coupling strength, as indicated by the anti-crossing frequency gap, is relatively large, reaching 350 MHz for the fundamental mode and slightly less for the quantized mode with two nodal lines along the Py width. These values position our ISRR design with a Py thin film among the systems with the largest photon-magnon anti-crossing frequency gap \cite{Bhoi2017}. We show that at frequencies around the center of these anti-crossing gaps, the SW modes are excited several times more efficiently than by the MSTL of the same dimensions and fed by the same current as the ISRR. Importantly, this enhancement is present in the weak coupling regime, which significantly reduces the resonator design requirements and the need to use low-loss materials. This finding opens up new perspectives and applications for hybrid photon-magnon systems and the use of resonant photon-magnon coupling for MW-to-SW transduction. An important advantage of using anti-gaps as antennas for spin-wave excitation is their robustness to impedance mismatch. While the reduction of MSTL sizes leads to impedance mismatch, the anti-gap ISRR supports effective excitation without the need for dimensional adjustments to the feed line.

\section{GEOMETRY AND METHODS}

\subsection{ISRR for SW excitation}

\begin{figure}
\centering
\includegraphics[width=\linewidth]{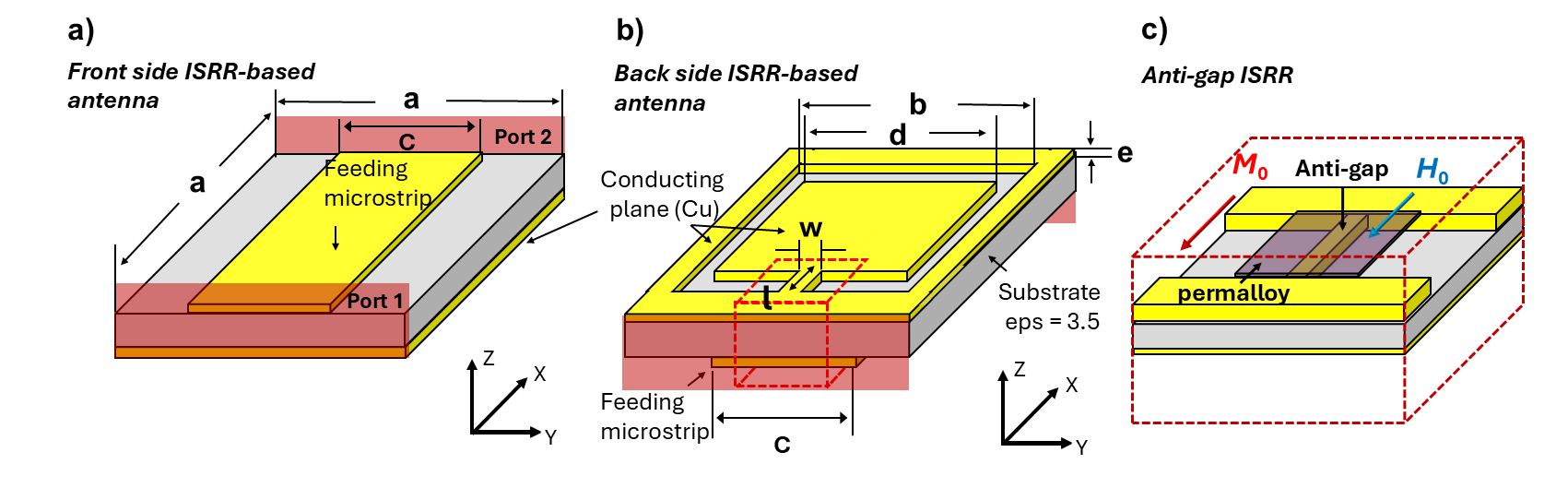}
    \caption{Schematic view of an ISRR used in the study: (a) view from the feeding line side, (b) view from the ISRR side, and (c) enlarged view of the anti-gap, over which a thin Py film is deposited. The Py film is saturated by the external bias magnetic field $H_0$ oriented along the $x$-axis.}
\label{fig:ISRR_1}
\end{figure}

\begin{figure}
\centering
\includegraphics[width=0.9\linewidth]{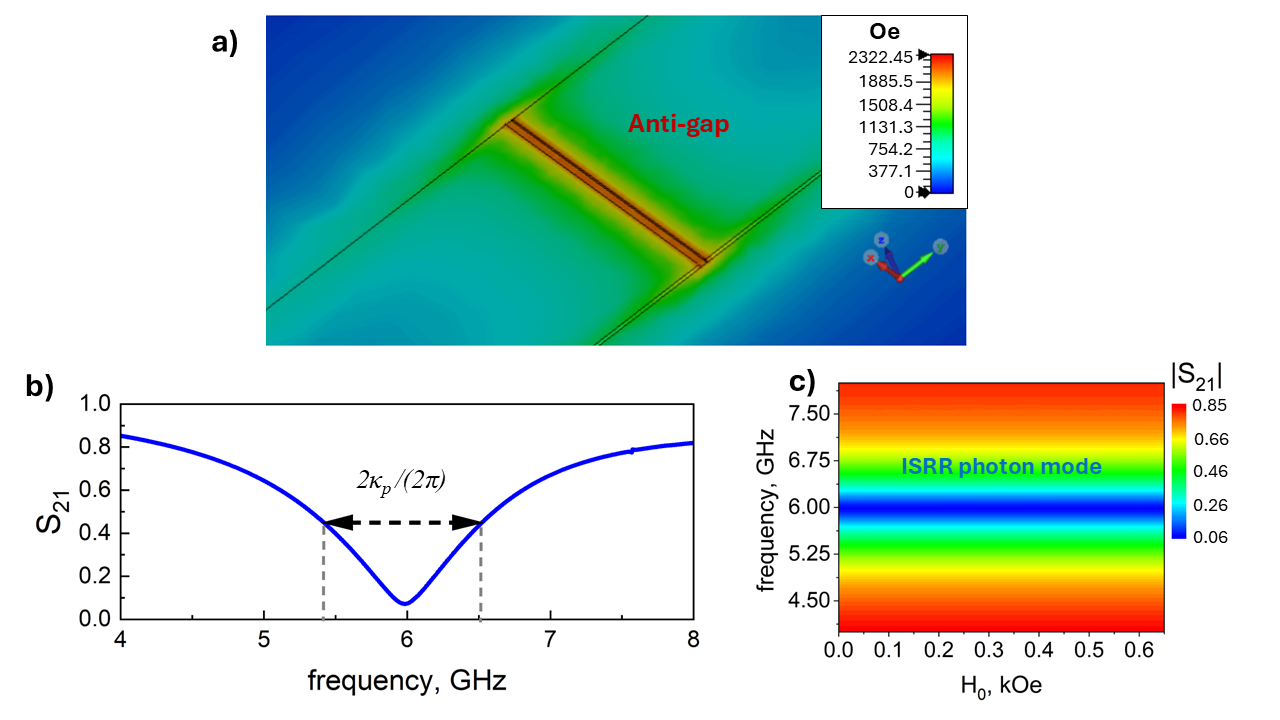}
    \caption{(a) The MW magnetic field distribution around the anti-gap of the ISRR. Results of CST simulations performed at 6~GHz and $H_0 = 0$ Oe. Most of the energy concentrates directly near the anti-gap strip, providing a relatively accurate amplification of the $h$-fields in this region. (b) Transmission coefficient ($|S_{21}|$) of the MW vs. frequency for the ISRR at $H_0 = 0$ Oe. (c) Color plot of $S_{21}$ in dependence on the frequency and bias magnetic field.}
\label{fig:ISRR_S}
\end{figure}

The proposed ISRR is shown in Fig.~\ref{fig:ISRR_1}.
A two-port microstrip line is used for feeding the ISRR with MW current [Fig.~\ref{fig:ISRR_1}(a)]. This feeding microstrip is positioned on the front side of a lossless dielectric substrate (quartz, $\varepsilon = 3.5$) with a thickness of 20 \(\mu\)m, while the ISRR is located on the opposite side of the substrate in a bottom ground plane [see Fig.~\ref{fig:ISRR_1}(b)].  The geometric parameters of the ISRR are selected in such a way that its resonance frequency is 6~GHz as shown in the transmission spectra ($S_{21}$ parameter)  in Fig.~\ref{fig:ISRR_S}(b) calculated through the feeding microstrip.  All geometric and material parameters of the ISRR are collected in Table 1 (for all metallic elements of the ISRR, we assume Cu with an electrical conductivity of $5.8 \times 10^7\ \text{S/m}$). 
The main interest in our investigation is associated with the anti-gap of the ISSR, which is a narrow metallic strip ($w$ is the anti-gap width and $l$ its length) connecting the general ground plane with a small metallic patch [see Fig.~\ref{fig:ISRR_1}(c)].

Photon-magnon coupling is most commonly studied with yttrium-iron-garnet (YIG) samples in the form of spheres or thin films~\cite{Tabuchi2014,Bhoi2017,Kaffash_2023,Bourcin2023}. This is due to a very narrow resonance linewidth—on the order of a single Oe. However, these high-quality YIG samples are typically grown on GGG substrates, which leads to increasing attenuation at low temperatures (linewidth above 40 Oe)\cite{Guo2023,Tabuchi2014}, limiting their potential usefulness in quantum technology. In addition, the fabrication of YIG is not compatible with CMOS technology, making it difficult to design integrated circuits with ferromagnetic and metallic elements in very close proximity. Thus, ferromagnetic layers are now being considered, and natural candidates are low damping transition metal alloys, e.g., Py\cite{Li2019,Hou_2019,Rincon2023,Inman2022} and FeCo \cite{Haygood2021}, which have a Gilbert damping constant on the order of $10^{-3}$. They have several times higher magnetization saturation, i.e., spin density, than YIG, and therefore larger coupling is also expected. Therefore, for our study we choose a thin Py film of finite extent [see the scheme in Fig.~\ref{fig:ISRR_1}(c)], i.e., its length is limited by the anti-gap length $l$. Its width and thickness will be varied in our study, but at the beginning we choose $w_{\text{Py}}=12$ \(\mu\)m and $t_{\text{Py}}=40$ nm.
We use standard parameters of Py in simulations. These are conductivity \( \sigma = 2.4 \) MS/m, saturation magnetization \( 4\pi M_s = 10600 \) Gs, Landé factor 2.0, and resonance linewidth \( \Delta H = 10 \) Oe~\cite{Zhao2016}.  
The external magnetic field \( H_0 \) is directed along the anti-gap strip, i.e., along the \(x\) axis [see Fig.~\ref{fig:ISRR_1}(c)], which allows SWs to be excited by the AC magnetic field and is the standard configuration used with MSTL~\cite{Mori_2022,Dmitriev1988,Vanatka_2021}.  

\subsection{Computational methods}

For the numerical simulations, we use the \textit{CST Studio Suite} package, which is well suited for the calculation of MW circuits and also offers the possibility of assigning customized constitutive parameters of the magnetic material. The response of the ferromagnetic material on the AC magnetic field is given by the Polder tensor~\cite{Polder1949VIIIOT}, which is well suited for magnetostatic SWs in ferromagnets of ellipsoidal shape, i.e., when exchange interactions and inhomogeneity of the demagnetizing field can be neglected\cite{Stancil2009}. In order to verify whether these approximations are justified for the Py film of finite extent proposed in our investigations, we have compared a selected set of CST results with the results of micromagnetic simulations with Mumax3~\cite{Vansteenkiste2014}, which include both effects.

One of the parameters to be considered in our study is the transmission coefficient, i.e. \( S_{21} \), which is usually the result of experimental studies of SW transmission and photon-magnon coupling measurements. The transmission spectra are calculated in the frequency range around the ISRR resonance, i.e., from 4 GHz to 8 GHz. We also calculate the spatial distribution of the AC \( h \) field inside and outside the ferromagnetic film. We also use open boundary conditions, whose implementation is provided in the solver.

\section{RESULTS AND DISCUSSION}   
   \subsection{The photon-magnon coupling}

In the following part of the study, we focus on the analysis of the transmission coefficient (\(S_{21} \) - parameter) as a function of frequency \(f\)and external magnetic field  \(H_0\). First, we consider this dependence for ISRR without a ferromagnetic film. The resulting \(|S_{21}|(f,H_0)\) spectrum [Fig.~\ref{fig:ISRR_S}(c)] reveals a single minimum at 6 GHz associated with the ISRR resonance [Fig.~\ref{fig:ISRR_S}(b)], and these ISRR resonance characteristics are independent from the external magnetic field [Fig.~\ref{fig:ISRR_S}(c)]. The full width at half maximum 2$\kappa_\text{p}/(2\pi)$ at the resonance is $1.28$~GHz.
The spatial magnetic field distribution at resonance at the anti-gap region, shown in Fig.~\ref{fig:ISRR_S}(a), clearly indicates that this is an active working part of the resonator around which the MW magnetic field ($h$) is enhanced. This part can act as an antenna to excite the magnetization dynamics and provide photon-magnon coupling in a ferromagnetic film if deposited in its vicinity, as we will demonstrate in the following part of the paper. 

\begin{figure}[h!]
    \centering
    \includegraphics[width=\textwidth]{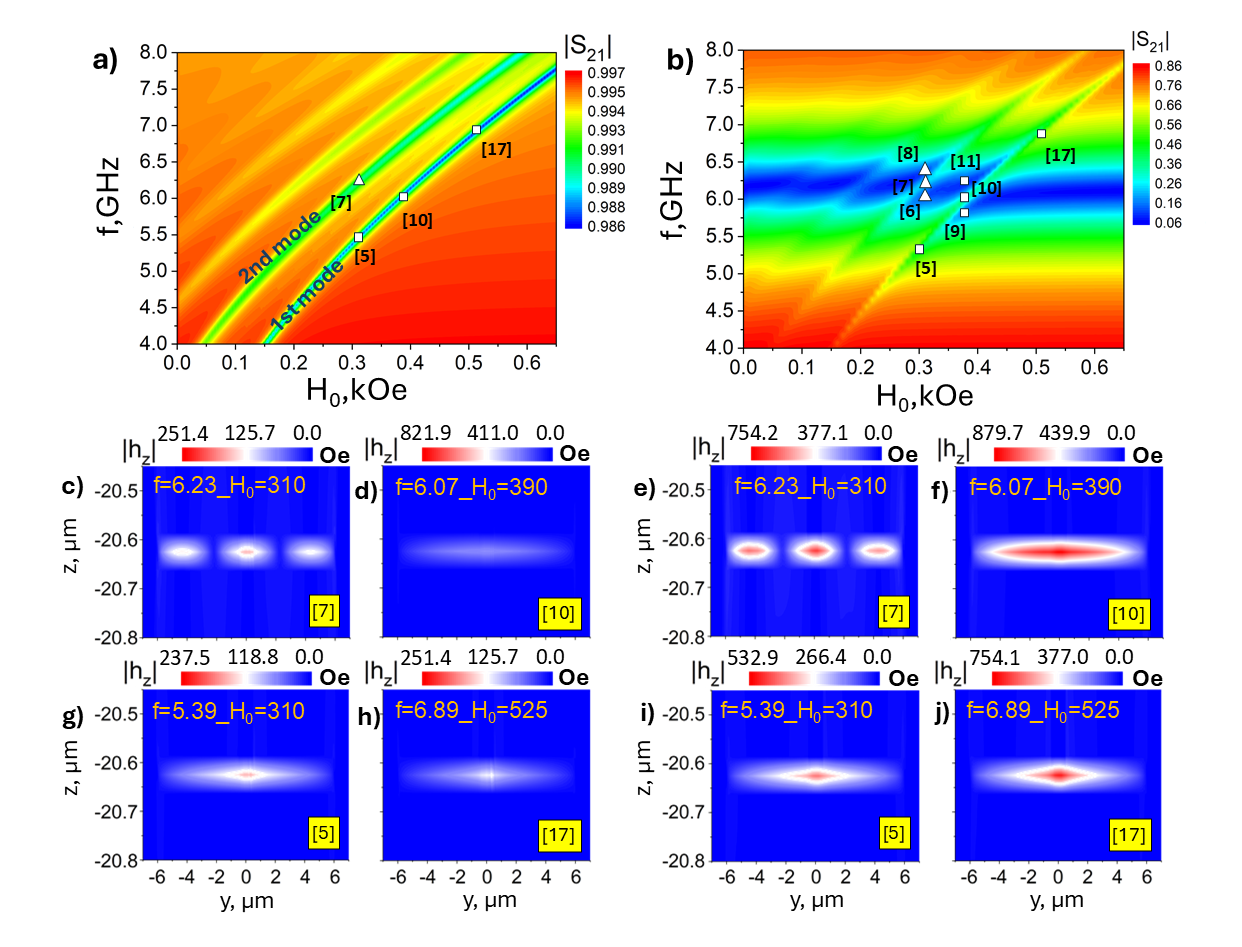} 
    \caption{MW transmission coefficient \(|S_{21}|\) versus frequency and static magnetic field for propagation through: (a) MSTL and (b) ISRR, both loaded with the same Py film. (c-j) Spatial distribution of the magnetic field component \(|h_z|\) in the $(y,z)$ cross section of the MSTL (c, d, g, h) and the ISRR anti-gap (e, f, i, j), crossing the Py film in the middle along the $x$ axis. The frequencies and magnetic fields for which the $|h_z|$ is plotted are marked in (a) and (b).}
\label{fig:3}  
\end{figure}

As a reference case, we consider a MSTL with the same dimensions as the anti-gap in the ISRR.
Just like in the ISRR configuration, the MSTL is placed on the front side of a lossless quartz substrate with a thickness of 20~\textmu m. A ground plane (Cu) with the same geometric dimensions as the substrate 6~mm $\times$ 6~mm and a thickness of 0.6~\textmu m is located on the opposite side.
Figure ~\ref{fig:3}(a) shows the \(|S_{21}|\) spectrum 
with 40 nm thick Py film (size of \(50 \times 12 \times 0.04\) \(\mu\)m) placed on top of it. The reduced transmission (blue and green color), generally by no more than 1\%, indicates energy absorption in the Py film, and conversion of the MW power into SWs. It shows the excitation of up to 5 width-quantized (along the $y$ axis) magnetostatic SW modes in the considered frequency range. The magnetic field distributions of the two lowest-order SWs are shown in Fig.~\ref{fig:3}(c-h). Due to the central positioning of the Py film with respect to the MSTL, only the symmetric modes along the $y$ axis are excited, e.g., fundamental (mode [10]) and mode with two nodal lines along the $y$ axis (mode [7]). 
In Fig.~\ref{fig:3}(a), we can also observe a slight decrease of the $|S_{21}|$ with increasing frequency. We attribute this effect to the increased reflection of the MW signal due to the increased impedance mismatch between the input port and the MSTL. This effect will be discussed later, in the presentation of the effectiveness of SW excitation.

The dependence of SW frequency on the magnetic field and SW amplitude distributions obtained with CST was verified by full micromagnetic simulations with Mumax3\cite{Leliaert2018FastMumax3}. In these simulations for the excitation of SWs in the Py film, we use the MW magnetic field spatial distribution extracted from the CST results [example is shown in Fig. S1(b) in the supplemental information (SI)]. We obtained perfect agreement for the first 4 bands of lowest frequency [see SI: Fig.~S1(a) in Sec. S1], which confirms the correctness of our approach used with CST.

Figure~\ref{fig:3}(b) illustrates the result of the simulation for ISSR loaded with the same Py film as in simulations with MTSL. 
We observe a superposition of both spectra, MW resonance in ISRR [Fig.~\ref{fig:ISRR_S}(c)] and SWs in Py film over MSTL [Fig.~\ref{fig:3}(a)]. However, the \(|S_{21}|\) spectrum also exhibits a specific feature called anti-crossings or mode repulsion effects. This feature is a marker of the presence of hybridization between MW photon mode and different SW modes, and thus there is sufficient photon-magnon coupling in an ISRR-Py system\cite{Tabuchi2014,ZARERAMESHTI2021}. 
At \( H_0 = 390 \) Oe, a distinct anti-crossing effect is observed, which is a signature of hybridization between the photon and the fundamental SW mode in Py. As the external magnetic field decreases, the higher-order SW modes of the Py film interact with the ISRR’s resonance mode, forming higher-order hybrid photon-magnon modes. In Fig.~\ref{fig:3}(b), we can identify up to 4 anti-crossings. This characteristic mode splitting, with only two hybrid branches, indicates rather independent coupling of the particular SW modes with the ISRR resonator, which is uncommon for the planar YIG samples with macroscopic dimensions\cite{Bhoi2017,Kaffash_2023}.  

The standard parameters describing the photon-magnon interaction are the coupling strength \(g \) and cooperativity \( C \), with which we can classify the type of photon-magnon coupling~\cite{Zhang2014,Kaffash_2023}.
Cooperativity is related to \(g \) by the following expression:
\begin{equation}
C = \frac{g^2}{\kappa_\text{p} \kappa_\text{m}},
\end{equation}
where \( \kappa_\text{m} \) is the half-linewidth at half-maximum of the magnon mode, \( \kappa_\text{p} \) is the half-linewidth at half-maximum of the photon mode. Both \( \kappa_\text{p} \) and \( \kappa_\text{m} \) are the values taken from the respective non-interacting systems.
In the framework of two coupled harmonic oscillators theory, the trajectory of two repulsed branches can be described as~\cite{BHOI20191}:
\begin{equation}
    \omega_{\pm} \approx \frac{1}{2} \left[ \left( \omega_\text{m} + \omega_\text{p} \right) \pm \sqrt{\left( \omega_\text{m} - \omega_\text{p} \right)^2 + \left( 2 g \right)^2} \right],
    \label{eq:placeholder}
\end{equation}
where \( \omega_\text{m} \) is the magnetic field dependent angular frequency of the particular SW mode in Py, \(\omega_\text{p} \) is the ISRR resonance angular frequency, both from the respective non-interacting systems.
The implementation of \(g \) in Eq.~(\ref{eq:placeholder}) gives us the possibility to measure its value directly from the \( |S_{21}|(f,H_0)\) graph, it is equal to half of the width of the gap between the branches measured on the frequency axis (anti-crossing frequency gap), at magnetic field where $\omega_\text{m} = \omega_\text{p}$.

The values of the photon-magnon coupling strength extracted from the simulation results are: \( g/(2\pi) \) = 175 MHz for the first magnon mode (mode [10] in Fig.~\ref{fig:3}(b)), \( g/(2\pi) \) = 165 MHz for the second magnon mode ([7] in Fig.~\ref{fig:3}(b)) and \( g/(2\pi) \) = 108 MHz for the third magnon mode. The half-linewidth at half-maximum of the magnon mode \( \kappa_\text{m}/(2\pi) \) is: 58 MHz for the first magnon mode ([10] in Fig.~\ref{fig:3}(b)), 88 MHz for the second magnon mode ([7] in Fig.~\ref{fig:3}(b)) and 128 MHz for the third magnon mode.  The corresponding cooperativity is \(C\)= 0.826 for the first magnon mode (mode [10] in Fig.~\ref{fig:3}(b)), \(C\)= 0.484 for the second magnon mode (mode [7] in Fig.~\ref{fig:3}(b)), and \(C\)= 0.143 for the third magnon mode. As expected, both $g$ and $C$ decrease with increasing order of the SW involved in the hybridization. According to standard criteria~\cite{Zhang2014,Kaffash_2023}, these values indicate that we are in a weak coupling regime due to \( g < \kappa_\text{p}\). In order to reach a strong coupling regime
(i.e. $C>1$ or $g>\kappa_\text{m}$, $g>\kappa_\text{p}$), the loss of the MW resonator must be reduced. This can be achieved by proper design of the ISRR geometry, e.g., by optimizing the coupling between the ISRR and the feed microstrip, or by using superconducting material instead of Cu, but here at the cost of lowering the operating temperature.
Nevertheless, the value of the obtained anti-crossing frequency gap (i.e., coupling strength) is among the largest exploited so far for photon-magnon coupling~\cite{Bhoi2017}. This includes MW cavities and split ring resonators as well as bulk YIG samples. Moreover, as we will show in the following part, the weak coupling regime is well suited to obtain enhanced efficiency of SW excitation.  

The presented results show that the anti-gap region acts as a confinement site for MWs, where the photon-magnon interaction is maximized when the resonance condition is fulfilled, allowing a spatially localized energy transfer from MWs into the magnonic system. Unlike conventional MSTL antennas, which predominantly excite pure magnetostatic waves and operate out of the MW resonance, the ISRR-based configuration facilitates the excitation of hybrid photon-magnon modes. 
The AC magnetic field distributions at different points of the $(f,H_0)$ plane are shown from the center of the Py film in Fig.~S2 in the SI, and from the perpendicular cross-section of the Py film in Fig.~\ref{fig:3}(e-f) and (i-j). It is clear that the magnetic field in the Py film is enhanced when it is taken from the line that follows the $f(H_0)$ dependence of the SWs. Along this line, the largest $|h_z|$ field is observed around the anti-crossing frequency gap (see modes [10] and [7] of the fundamental and quantized SWs, respectively). It indicates that the hybrid photon-magnon systems may also have other usefulness that has not yet been explored in the literature, i.e., the excitation of SWs.

\subsection{Excitation of SWs by ISRR and comparison with MTSL}

 Interestingly, the anti-crossing frequency gap is detectable in transmission or reflection measurements, but when the magnetization oscillation is measured directly, e.g.,  with Brillouin light scattering spectroscopy, only the SW branch, which is unaffected by the coupling, is detected\cite{Kaffash_2023}. 
In addition, it has recently been shown that photon-magnon coupling, in its weak coupling regime, can be used to excite the short-wavelength exchange standing SWs formed across the thickness. This has been demonstrated for a bilayer ferromagnet composed of Py (30 nm thick) and YIG (micrometer thick) with millimeter lateral dimensions. All this suggests that photon-magnon coupling may be useful to excite SW in thin ferromagnetic films, and that even weak coupling may be advantageous for this purpose.

\begin{figure}[h!]
    \centering
    \includegraphics[width=\textwidth]{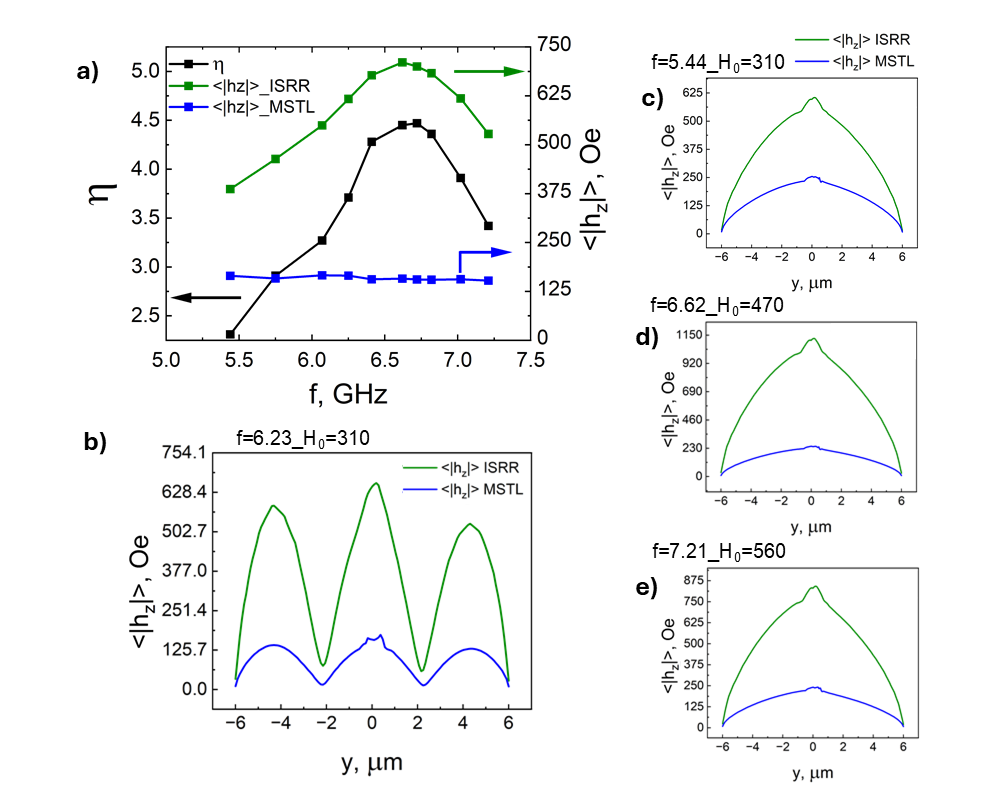} 
    \caption{(a) Enhancement ($\eta$) of SW excitation by ISRR versus MSTL as a function of SW frequency - black dots. Green and blue points show the average AC magnetic field in the Py film for SWs excited by ISRR and MTLS, respectively.  (b-e) AC \(|h_z|\) field at the centre line (along the \(y\) axis) of the Py film for the: (b) the first quantized SW mode and (c-e) the fundamental SW mode at 5.44, 6.62 and 7.21 GHz.}
\label{fig:MSTL-ISRR}  
\end{figure}

To verify this hypothesis, we perform a comparative evaluation of the SW mode excitation effectiveness for two different antennas: the MSTL and the ISRR. The cross-section of the MSTL and ISRR anti-gap 1.0 × 0.6 \(\mu\)m is identical for both structures.  We focus on the comparison of the $z$ component of the AC magnetic field (\( h_z \)) strength in the thin Py film, since this field is a direct result of the SW dynamics in the Py film and can be obtained directly from CST Studio Suite. The Py film with dimensions of 50 × 12 × 0.04 \(\mu\)m, the same as in the previous section, is placed centrally on an MSTL antenna in the first case and on the ISRR anti-gap in the second. For accurate comparison between MSTL and ISRR configurations, the current at the emitting port is set to 1~mA in both cases. The results [e.g., see Fig.~\ref{fig:3}(c)-(d) for the MSTL and (e)-(f) for the ISRR] show that the same SW modes appear in both configurations within the same external field/frequency range. However, the intensities of the induced AC magnetic field in Py are different.
This is visualised in Fig.~\ref{fig:MSTL-ISRR}(c)-(e), which shows $|h_z|$ in the center of the Py film, along the $y$ axis. The profiles are shown for frequency and magnetic field values following the SW line of the fundamental mode [Fig.~\ref{fig:3}(a)], for both ISRR (green line) and MSTL (blue line). The AC magnetic field induced by the SW in the Py film placed on the anti-gap of the ISRR (maximum value reaching 1000 Oe) is several times stronger than over the MSTL (with a maximum value not exceeding 260~Oe).
Fig.~\ref{fig:MSTL-ISRR}(b) shows the AC field profile for point [7] in Fig.~\ref{fig:3}(a), i.e., for the first width-quantized SW mode. The maximum amplitude is about 630~Oe and 130~Oe for ISRR and MSTL, respectively, thus also demonstrating the enhancement of the SW by the ISRR for higher-order SW modes as well.

The values of the $h_z$ amplitude averaged along the sample width ($<|h_z|>$) as a function of SW frequency are shown in Fig.~\ref{fig:MSTL-ISRR}(a) for ISRR and MSTL with green and blue colored dots respectively. As expected, in the system with MSTL excitation, the average $h_z$ field depends only weakly on the frequency of the MWs and remains at values below 240~Oe. However, for the ISRR, the average AC magnetic field increases from about 400 Oe at 5.44 GHz and reaches a maximum above 700 Oe at 6.62 GHz. Unexpectedly, this maximum is not at the MW resonance frequency, i.e. 6~GHz, but is shifted up to 6.62~GHz.

To quantify the enhancement of the SW excitation efficiency in ISRR compared to MSTL, we introduce the enhancement factor $\eta$, which is defined as the ratio of the spatial averages of the $|h_z|$ field along the $y$ axis inside the Py film in ISRR and MSTL systems: $\eta=\frac{<h_z^{\text{ISRR}}>}{<h_z^{\text{MSTL}}>}$. The results are shown in Fig.~\ref{fig:MSTL-ISRR}(a) with black dots, and this dependence is qualitatively similar to $<|h_z|>(f)$ for the ISRR. The enhancement starts at frequencies below ISRR resonance, at 5.44 GHz $\eta = 2.31$, it increases with increasing frequency, reaching a maximum $\eta$ above 4.45 at 6.62~GHz, then at higher frequencies the enhancement decreases. These results clearly show an enhancement of SW excitation several times with ISRR compared to MSTL. This is an effect of the MW resonance of the ISRR, in which enhanced concentration of the MW magnetic field is obtained near the anti-gap.
However, the maximum enhancement shifts from the photon-magnon resonance condition to higher frequencies, i.e. from 6~GHz to 6.62 GHz. We have not found a clear explanation for this effect. It seems that a similar effect has been observed in the Brillouin light scattering measurements mentioned above\cite{Kaffash_2023}. It may be related to the effect of non-reciprocity combined with the negative permeability introduced into the ISRR spectra by the ferromagnetic film, as has recently been demonstrated for a similar structure but loaded with the macroscopic YIG film and placed on the side of the fed microstrip line\cite{Kim2024}. However, the verification of this hypothesis requires further in-depth investigations.
   
   \subsection{Influence of the thickness of the ferromagnetic film}

We have shown that in the weak coupling regime, an ISRR effectively excites the SWs in a nanometer-thick Py film at frequencies corresponding to magnetostatic SWs and not to hybridized photon-magnon modes. These frequencies thus cross the center of the anti-crossing frequency gap. An interesting question is whether the excitation efficiency depends on the photon-magnon coupling strength. The natural way to increase the coupling strength is to increase the volume of the ferromagnet involved in the coupling\cite{Wagle_2024}. We do this by increasing the thickness of the Py film, keeping its lateral dimensions the same, that is $w_\text{Py}=12$ \(\mu\)m and $l=50$ \(\mu\)m, and using the same ISRR and MSTL for the excitation.

\begin{figure}[h]
    \centering
    \includegraphics[width=\textwidth]{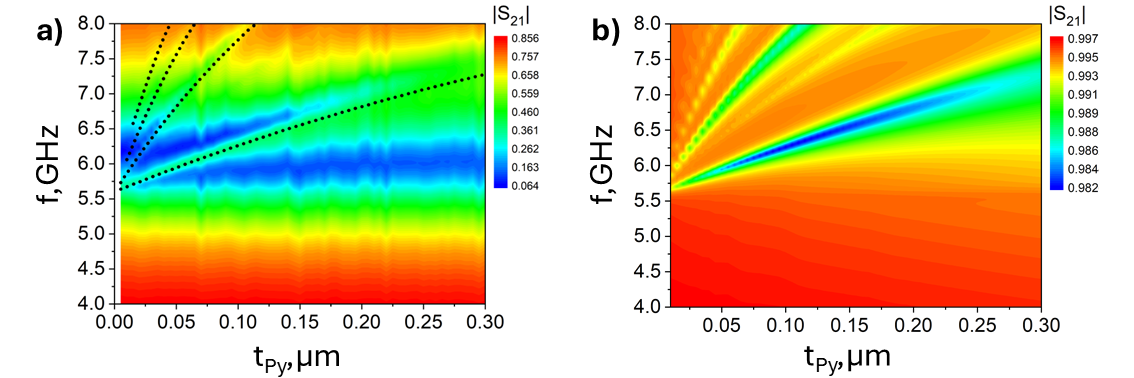} 
    \caption{Transmission coefficient \(|S_{21}|\) as a function of frequency and Py film thickness at fixed external magnetic field \(H_0 = 365\) Oe for (a) ISRR and (b) MSTL. The black dashed lines in (a) indicate the frequencies of the SW modes extracted from (b).}
\label{fig:spectra-tp}    
\end{figure}

First, we show the $|S_{21}|$ as a function of frequency and Py film thickness \(t_{Py}\) at \(H_0 = 365\) Oe for the case where the MSTL acts as a SW antenna in Fig.~\ref{fig:spectra-tp}(b) (these data from CST were confirmed in micromagnetic simulations). The frequencies of all SW modes increase with \(t_{Py}\), since for magnetostatic waves the frequency is dependent on $k t_{Py}$\cite{Stancil2009}, where $k$ is a wave number associated with a particular SW mode. In our case $k$ is determined by the film width, although in ferromagnetic extended films the frequency of the fundamental SW mode (i.e., the Kittel mode) does not depend on $t_\text{Py}$, here the amplitude of the SWs is pinned along the $y$-axis [see Fig.~\ref{fig:3}(f), (i) and (j)], due to the dipolar interactions\cite{Guslienko2003}. Thus, in Fig.~\ref{fig:spectra-tp}(b) we observe a monotonic increase of the frequency also for the fundamental mode. 
 
An analogous dependence, but for the Py film on ISRR, is shown in Fig.~\ref{fig:spectra-tp}(a). Qualitatively, it looks very similar, but the spectra are dominated by the ISRR MW resonance, indicated by the horizontal line at 6 GHz. The lines of the magnetostatic SWs from Fig.~\ref{fig:spectra-tp}(b) are marked by the black dashed lines. They fit very well with the anti-crossing frequency gaps in the photon-magnon coupling region and follow the enhanced transmission at high frequency ($f>7.5$ GHz).
The increasing splitting of the MW resonant mode near the anti-crossing with the fundamental mode suggests an increase of the anti-crossing frequency gap. 
Indeed, the coupling strength increases to $g/2\pi=267$ MHz for the $t_\text{Py}=100$~nm, as shown in the $S_{21}$ dependence on $f$ and $H_0$ in Fig.~S4(a) in SI.

In order to obtain the effectiveness of the SW excitation in terms of perceptible photon-magnon coupling, we have carried out comparative simulations of the MSTL and the ISRR loaded with the 100 nm thick Py film (lateral dimensions remain the same as before). The AC magnetic field distribution in the Py film at the center of the anti-crossing frequency gap is shown in Fig.~S4(a) in SI, together with the $h_z$ field excited with the MSTL. From this, we found that the enhancement of the SW excitation by the ISRR with respect to the MSTL is $\eta =2.8$, which is weaker than that found in the 40~nm Py film.
This indicates that the weak coupling regime is preferred for SW excitation, and the optimal value of $g$ depends on the system under consideration and can be optimized.

\section{CONCLUSIONS}

In this work, we have addressed a key challenge in magnonics, i.e., the efficiency of MW signal conversion into SWs, by using a planar MW resonator implemented as an ISRR for the local excitation of SWs in a nanometer-thin ferromagnetic metal film.
We have shown with full electromagnetic numerical simulations that the ISRR structure allows simultaneously high spatial localisation and efficient MW energy transfer into the magnonic system in a small area of the ISRR anti-gap, confining the interaction region to deep MW subwavelength dimensions below \(1 \times 12 \) \(\mu\)m at the 6 GHz operating frequency, while maintaining the resonant enhancement.
This creates a suitable condition for photon-magnon coupling even in a nanometer-thin metallic ferromagnet, with an anti-crossing frequency gap exceeding 350~MHz, which is among the largest values obtained so far, even compared to bulk ferromagnets. Although it is in a weak-coupling regime due to the broad MW resonance linewidth of the ISRR, we have achieved a significant improvement in the efficiency of MW-to-SW conversion in the nanometer-thick Py film compared to conventional MSTL, approximately 4.2 times for the fundamental SW mode and 4.0 times for the width-quantized SW mode. Unlike conventional MSTL antennas, which predominantly excite only pure magnetostatic waves, the ISRR-based antenna simultaneously excites both pure magnetostatic waves and hybrid photon-magnon modes, enhancing the efficiency of MW-to-SW conversion.

The features presented make the proposed device a promising solution for scalable and high performance magnonic circuits. This is because the only element tuned for SW excitation is the anti-gap line, while the feed line and the other parts of the ISRR can remain macroscopic in size and can be easily integrated into the MW circuits. This overcomes the limitations of conventional planar MW antennas and paves the way for the next generation of magnonic MW signal processing devices with enhanced integration capabilities. Future research will focus on optimization of the ISRR, deep understanding of the non-reciprocity suggested by our results, and experimental validation and further optimization of the antenna geometry to extend its applicability in nanoscale magnonic circuits or to achieve strong photon-magnon coupling with the sub-micron metallic ferromagnet, which can be useful for quantum technologies.

\section{ADDITIONAL INFORMATION}

\subsection{Micromagnetic modeling with MuMax3}

For the comparative analysis of the spectrum of magnetostatic modes, a simulation was performed in the MuMax3 micromagnetic simulation package~\cite{Vansteenkiste2014}. 
During the simulation, the ferromagnetic Py film was represented by a parallelepiped with dimensions along the \(x, y\) and \(z \) axes as \(50 \times 12 \times 0.040\) \(\mu\)m. Since only long-wavelength magnetostatic modes were considered during the simulation, the simulation region was divided into cells with dimensions \((x, y, z)\) of \(48.828 \times 46.875 \times 0.040\) nm. The external static field was applied along the \(x \)-axis, and the alternating high-frequency field was applied along the \(y \)-axis.
The alternating high-frequency field was set as a function 
\[
B_{\text{ext\_AC}}(x,y,z,t) = G(x,y,z) \cdot F(t)
\]
where \( G(x,y,z) \) is the spatial distribution of the components of the vector \( B_{\text{ext\_AC}} \), and \( F(t) \) is the time dependence of its components.  

The spatial distribution of the components of the vector \( B_{\text{ext\_AC}} \), i.e., the function \( G(x,y,z) \), was calculated from results previously obtained in CST Studio Suite and represents the spatial distribution of the normalized amplitudes of the  magnetic components of the electromagnetic field \( h(x,y,z) \) in the plane above the anti-gap of the ISRR. 
The function \( F(t) \) was given in the form:  
\[
F(t) = \text{sinc}(2\pi f (t - t_0)),
\]  

where \(f\) - frequency and equal 10 GHz in simulation, \(t_0\) - position of the center of \(sinc\)-function on time axis. 
The magnetic parameters of Py, such as the Landé factor \( g  \),  the saturated magnetization \( M_S \) and the damping factor \( \alpha \) were used the same as in the simulation in CST Studio Suite.

\subsection{Geometric parameters of the ISRR}

\begin{table} 
  \caption{Geometric parameters of the ISRR used in the paper and shown in Fig.~\ref{fig:ISRR_1}.}
  \label{tbl:example}
  \centering
  \begin{tabular}{|c|c|c|}
    \hline
    Dimensions  & \(\mu\)m & Description \\  
    \hline
    $a$ & 6000 & size of substrate \\  
    $b$ & 4000 & outer size of ISRR \\  
    $c$ & 20 & width of feeding stripline \\  
    $d$ & 3900 & inner size of ISRR \\  
    $e$ & 0.6 &Ground, MSTL, metallic strip (thickness)\\
    $w$ & 1 & width of ISRR anti-gap \\  
    $l$ & 50 & length of ISRR anti-gap \\   
    \hline
  \end{tabular}
\end{table}

\begin{acknowledgement}
The research has received financial support from the National Science Center of Poland, OPUS-LAP grant no. 
2020/39/I/ST3/02413. 

\end{acknowledgement}

\begin{suppinfo}

The Supporting Information provides additional information concerning the verification of the numerical calculations from CST Studio Suite using micromagnetic modelling in the package MuMax3. The Supporting Information also includes the |S21| spectrum and AC magnetic field spatial distribution for the ISRR, the dependence on the Py film width, and a comparison between the excitation of SWs by ISRR and MTSL for a film thickness of 100 nm.

\end{suppinfo}

\bibliography{acs-achemso}

\end{document}


\maketitle

\begin{abstract}
One of the main challenges in magnonics is the efficiency of the conversion of microwave signals into spin waves. This efficiency is low due to the significant mismatch between microwave and spin wave wavelengths in the GHz range $10^{-2}$ m and $10^{-8}$ m, respectively, leading to high energy consumption in magnonic circuits. To address this issue, we propose an approach based on a planar inverse split-ring resonator (ISRR) loaded with a nanometer-thick Py film and exploiting the photon-magnon coupling effect. Our numerical studies show that the ISRR-based antenna achieves more than a fourfold improvement in conversion efficiency compared to a conventional single microstrip transmission line at frequencies and bias magnetic fields around the anti-crossing frequency gap. This has been demonstrated in the weak photon-magnon coupling regime for the nanometer-thin permalloy film with micrometer lateral dimensions. Further optimization of the ISRR can help to achieve the strong coupling regime, making the system potentially useful for quantum technology.
   Our compact and efficient antenna design offers a significant advantage over standard microstrip lines, paving the way for scalable and powerful magnonic circuits for microwave signal processing.
\end{abstract}

\section{Verification using micromagnetic modelling in the package MuMax3}

\begin{figure}[h]
    \centering
    \includegraphics[width=\textwidth]{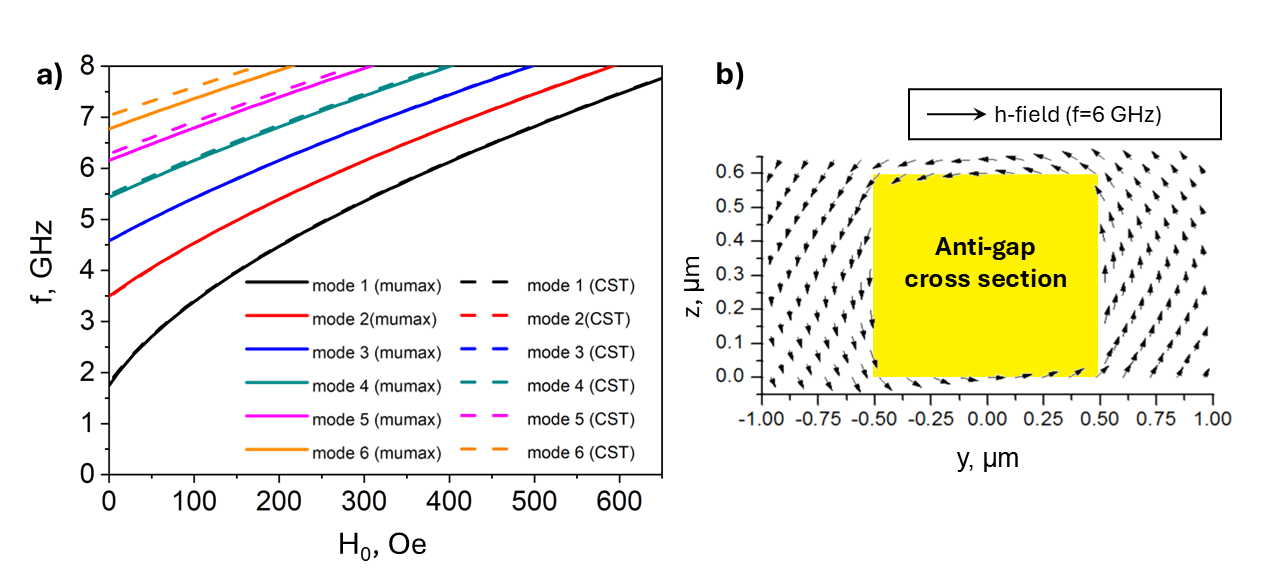} 
    \caption{Numerically calculated magnetostatic modes frequencies on external magnetic field strength for a MSTL (a). Solid lines (MuMax3) and dashed lines (CST Studio Suite); The MW \(h \)-field spatial distribution near the ISRR's anti-gap (cross-section shown by yellow on graph) at the resonance frequency of 6 GHz (\(H_0 = 0\) Oe) (b).}
 \label{fig:spectra-mumax}   
\end{figure}

Figure~S\ref{fig:spectra-mumax}(a) shows a comparison of the frequency of SWs vs. external magnetic field in 40 nm thick Py film deposited on the MSTL (microstrip transmission line), and calculated with CST (dashed lines) and Mumax3 (solid lines). Very good agreement was reached with the fourth SW band in the considered frequency range. The differences observed in higher-order SW modes are due to exchange interactions neglected in CST. In Fig.~S\ref{fig:spectra-mumax}(b) shows the MW \(h \)-field spatial distribution in the $(y,z)$ cross section emitted by the ISRR's anti-gap. This field spatial distribution was calculated in CST and then used in micromagnetic simulations with Mumax3 as the excitation MW field.

\section{The \(|S_{21}|\) spectrum and AC magnetic field spatial distribution}

\begin{figure}[h]
    \centering
    \includegraphics[width=\textwidth]{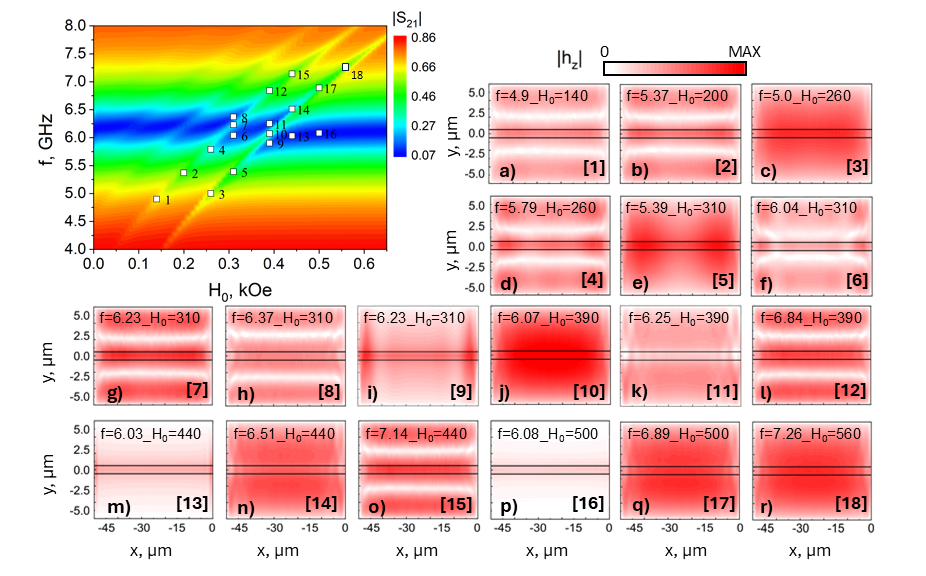} 
    \caption{MW transmission coefficient \(|S_{21}|\) versus frequency and static magnetic field for the ISRR (main graph), and spatial distribution of the AC magnetic field component \(|h_z|\) in the $(x, y)$ cross section in 50 × 12 × 0.04 \(\mu\)m Py film for the corresponding points on Fig.~S\ref{fig:spectra-spat_distribution} main graph (a-r).}
 \label{fig:spectra-spat_distribution}
\end{figure}

Fig.~S\ref{fig:spectra-spat_distribution} shows the MW transmission coefficient \(|S_{21}|\) versus frequency and static magnetic field  for the SWs excited by ISRR  (main graph). The points of interest are marked on the graph and AC magnetic field component \( |h_z | \) spatial distribution is analyzed for these points.

Points 3, 5, 10, 14, 17, 18 show a uniform field distribution which corresponds to a pure fundamental SW mode. Points 9, 11 which lie on a hybridized part of the photon-magnon spectrum also demonstrate quasi-uniform AC \(h\)-field character. Comparison of red color intensity of the points 3, 5, and 10 indicate a stronger \(h\)-field for point 10, which is confirmation of the positive influence of mode hybridization on the efficiency of SW excitation. The changes of field distribution from point 9 to 13 and 16 demonstrate the transformation of the AC \(h\)-field from quasi-uniform (fundamental SW mode in non-infinite ferromagnetic sample) to high localized that is typical for a microstrip line.

Similar features are seen for points corresponding to the second SW mode: points 1, 2, 4, 6-8, 12, 15. The spatial distribution has two nodes along the y-axes that indicate the highest order of this mode.  The more deep red color (point 7) is also in the middle of the anti-crossing area.

\section{Dependence on the Py film width}

\begin{figure}[h]
    \centering
    \includegraphics[width=\textwidth]{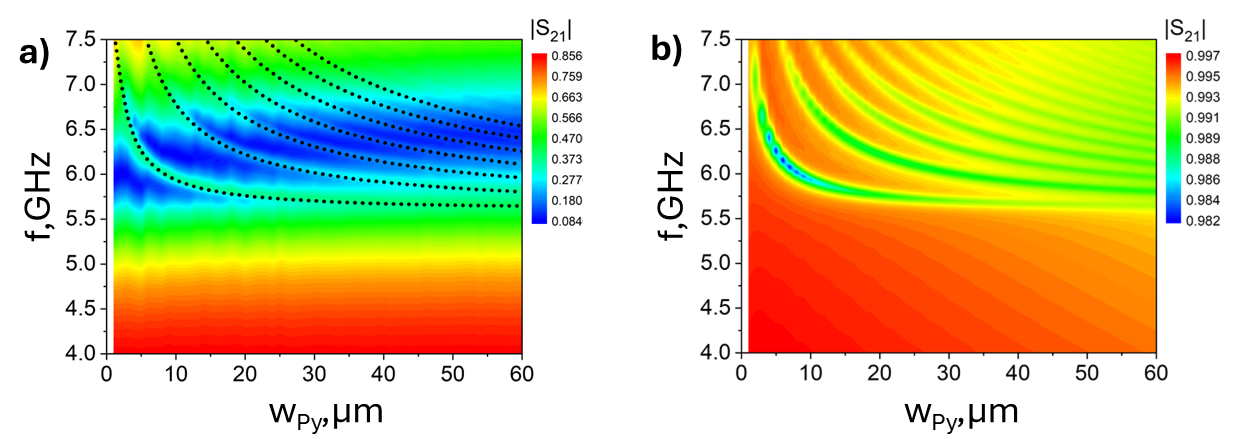} 
    \caption{\(|S_{21}|\) spectra vs Py film width at \(H_0 = 365\) Oe for the ISRR SWs excitation (a); MSTL SW excitation (b).}
 \label{fig:spectra-w}   
\end{figure}

Figure~S\ref{fig:spectra-w} illustrates the numerically calculated 2D \(|S_{21}|\) spectra as a function of Py film width \(w\) at a specific external magnetic field strength of \(H_0 = 365\) Oe for two cases:  
(a) an anti-gap of ISRR with a Py film placed on it, and  
(b) a MSTL with a Py film.  

Both Fig.~S\ref{fig:spectra-w}(a) and S\ref{fig:spectra-w}(b) were obtained using the CST Studio Suite package, whereas the FT operation result on the time snapshots of spatial distribution of excited magnetization in the Py film of different width (Figure ~\ref{fig:spectra-w}, c) was calculated using MuMax3.
As we can see from Fig.~S\ref{fig:spectra-w} the sharp decreasing of SW modes frequencies on Py film width growing.

In Fig.~S\ref{fig:spectra-w} we have shown photon-magnon hybridization with up to three spin wave modes, which have different quantizations along the Py width. Increasing the width of the ferromagnet will increase the wavelength (as we can associate $k \propto 1/w$) and consequently decrease the frequency. Indeed, such a dependence is found in the spectra of SW excited by the MSTL for Py films of different widths, Fig.~S\ref{fig:spectra-w}(b). Interestingly, the analogous spectra obtained in the ISRR look different, as shown in Fig.~S\ref{fig:spectra-w}(a). However, there is a general trend in the lines associated with the SWs (green areas intersecting the blue). In addition, the resonance frequency of the ISRR increases with $w$, which may be the result of changing of the upper anti-gap volume. The dielectric constant $\varepsilon$ of this volume increased from 1 (when ferromagnetic sample is absent) to its highest value when the sample size increased.

\section{Excitation of SWs by ISRR and comparison with MTSL for 100 nm film thickness}

  \begin{figure}[h]
    \centering
    \includegraphics[width=\textwidth]{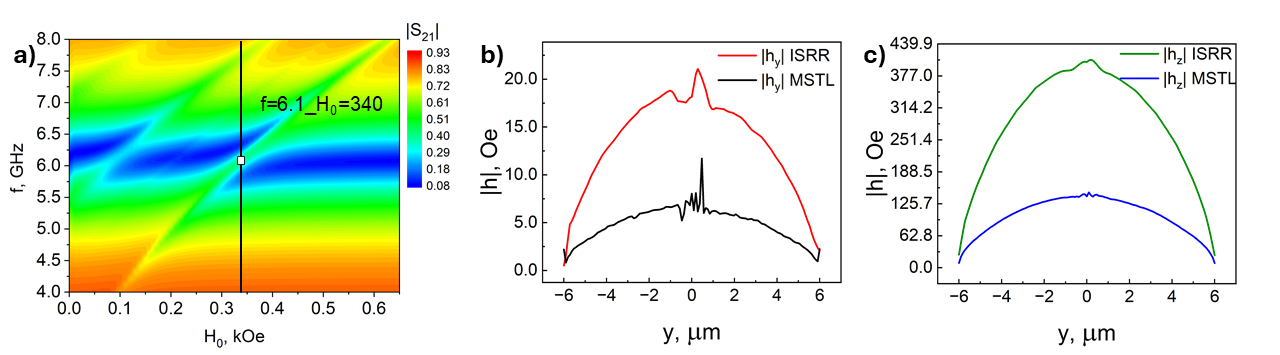} 
    \caption{\(|S_{21}|\) spectra vs Py film width at \(H_0 = 365\) Oe for the ISRR SW excitation (a); MSTL SW excitation (b).}
 \label{fig:spectra-100nm}   
\end{figure}